\documentclass[final,5p,times,twocolumn]{elsarticle}

\usepackage{amssymb}
\usepackage{array}
\usepackage{pifont}
\usepackage[colorlinks=true, linkcolor=blue, citecolor=blue, urlcolor=cyan]{hyperref}

\journal{}

\begin{document}

\begin{frontmatter}

\title{Large Language Model-Driven Database for Thermoelectric Materials\tnoteref{1}}

\author[a]{Suman Itani}
\ead{suman.itani@unh.edu}
\author[a,b]{Yibo Zhang}

\author[a]{Jiadong Zang}
\ead{jiadong.zang@unh.edu}

\affiliation[a]{organization={Department of Physics and Astronomy, University of New Hampshire},
            addressline={9 Library Way}, 
            city={Durham},
            postcode={03824}, 
            state={NH},
            country={USA}}

\affiliation[b]{organization={Department of Chemistry, University of New Hampshire},
            addressline={23 Academic Way}, 
            city={Durham},
            postcode={03824}, 
            state={NH},
            country={USA}}

\begin{abstract}

Thermoelectric materials provide a sustainable way to convert waste heat into electricity. However, data-driven discovery and optimization of these materials are challenging because of a lack of a reliable database. Here
we developed a comprehensive database of 7,123 thermoelectric compounds, containing key information such as chemical composition, structural detail, seebeck coefficient, electrical and thermal conductivity, power factor, and figure of merit (ZT). We used the GPTArticleExtractor workflow, powered by large language models (LLM), to extract and curate data automatically from the scientific literature published in Elsevier journals. This process enabled the creation of a structured database that addresses the challenges of manual data collection. The open access database could stimulate data-driven research and advance thermoelectric material analysis and discovery.

\end{abstract}




\end{frontmatter}


\section{Introduction}
\label{Introduction}

The world's growing energy challenges and environmental concerns have put sustainable energy solutions in the spotlight. Thermoelectric materials have emerged as critical components in promising technologies that directly convert heat into electricity and vice versa \cite{disalvo1999thermoelectric,bell2008cooling}.
These materials facilitate direct conversion between thermal and electrical energy via the Seebeck and Peltier effects \cite{rowe2018thermoelectrics, snyder2008complex}. This capability provides significant advantages in many applications, such as waste heat recovery, solid-state refrigeration, and thermal management in microelectronics \cite{disalvo1999thermoelectric, wang2024realizing, siddique2023phase,rowe1999thermoelectrics}. Despite their potential, the broader use of thermoelectric technologies has been limited by their relatively low energy conversion efficiency \cite{zhang2015thermoelectric}. As a result, their practical applications have largely remained confined to specialized areas. Thus, there is a pressing need to discover new thermoelectric materials.

The performance of these materials is typically quantified by the dimensionless figure of merit ($ZT$) which is defined as $$ ZT = \frac{S^2 \sigma T}{\kappa}$$ where S is the Seebeck coefficient, $\sigma$ is the electrical conductivity, T is the absolute temperature, and $\kappa$ is the thermal conductivity \cite{wood1988materials}. A high ZT value necessitates materials with high thermopower, high electrical conductivity, and low thermal conductivity. There has seen steady improvements of ZT value through careful engineering of electronic and phononic properties \cite{ma2021review}. However, the complex interplay between electrical conductivity, Seebeck coefficient, and thermal conductivity presents a fundamental challenge in material design, as these properties are often interdependent and must be optimized simultaneously \cite{feng2018advanced,ye2017ultra}.

Historically, the development of thermoelectric materials relied heavily on experimental trial and error, combined with traditional theoretical simulations, such as density functional theory (DFT) and molecular dynamics (MD) \cite{li2022machine, namsani2017interaction, jain2016computational}. These methods are effective for specific studies but require significant time and computational resources. This makes it difficult to apply them to the vast range of potential thermoelectric compounds. High-throughput computational approaches have been developed to accelerate this process. These methods facilitate systematic screening of materials \cite{gorai2017computationally, deng2024high, sarikurt2020high}. However, for many complex materials or doped systems, the computational cost remains prohibitively high \cite{chelikowsky2011computational}. This bottleneck has driven a shift toward data-driven approaches and machine learning (ML). These methods use existing data to quickly predict thermoelectric properties, bypassing the computational demands of first-principles calculations. \cite{jia2022unsupervised,wang2020machine, wang2023critical,antunes2023predicting}.

To effectively implement machine learning (ML) for thermoelectric material discovery, a large and high-quality dataset of thermoelectric properties is indispensable. Several publicly available databases \cite{ricci2017ab, choudhary2020joint, yao2021materials, gorai2016te, wang2011assessing, carrete2014finding, xi2018discovery, fang2024wenzhou} provide information on thermoelectric materials and their associated properties; however, the majority are derived from first-principles calculations and are often limited to ideal, undoped crystalline structures, which do not fully capture the complex behavior of real-world thermoelectric materials. Experimental databases \cite{na2022public,katsura2019data,gaultois2013data, chen2019machine, priya2021accelerated}, in contrast, are often small in size and manually curated. Furthermore, most existing data sets are limited in scope, typically focusing on one or few properties , rather than encompassing a comprehensive set of properties necessary for designing novel thermoelectric materials.

To address these limitations, Sierepeklis O. and Cole J.M. developed a database by employing ChemDataExtractor \cite{swain2016chemdataextractor}, a natural language processing (NLP) based tool, to automate the extraction of thermoelectric data from scientific literature. Through this method, they compiled a data set consisting of approximately 10,641 unique thermoelectric compounds, enriched with a diverse range of features \cite{sierepeklis2022thermoelectric}. This innovative work marks a significant advancement in automating the collection of thermoelectric properties from large volumes of unstructured textual data. Nevertheless, challenges persist. The primary concern is the reduced accuracy of data extraction when articles describe multiple compounds along with their respective properties, which can compromise the overall reliability of the database. Furthermore, this database combines experimental and computational data without clear differentiation, limiting its utility for specific applications. 

A notable gap in existing thermoelectric materials databases is the lack of structural properties, which are essential for understanding the intrinsic relationships between material structure and thermoelectric performance. Addressing this gap is critical for developing comprehensive datasets that can enable advanced machine learning-based predictive modeling of thermoelectric materials.

Large Language Models like GPT have become powerful tools for developing comprehensive scientific databases. In a recent advancement by the authors, Zhang et al. created the GPTArticelExtractor \cite{zhang2024gptarticleextractor}—an innovative workflow designed to automatically extract structured information from scientific literature.
Using this approach, they constructed a highly accurate large-scale database of magnetic materials and built machine learning models to identify promising high-performance magnetic compounds. Their work highlights the potential of AI-driven techniques for managing scientific research information and accelerating material discovery \cite{itani2024northeast}.

In this work, we developed a comprehensive database of 7,123 thermoelectric compounds, incorporating detailed thermoelectric properties and structural information using the GPTArticleExtractor. The database includes key parameters such as chemical composition, Seebeck coefficient, electrical and thermal conductivity, power factor, figure of merit (ZT), crystal structure, lattice structure, lattice parameters, and space group. Thermoelectric properties are provided with their corresponding measurement temperatures. Using LLMs, we automated the extraction and organization of data from scientific literature, addressing the limitations of manual data collection. The resulting structured dataset serves as a valuable resource for data-driven research and machine learning applications, supporting the study and design of thermoelectric materials. By enhancing access to comprehensive material data, this database provides a foundation for future efforts in optimizing thermoelectric properties and advancing energy-efficient technologies.

\section{Material and methods}
\subsection{DOIs Collection}
To begin our study, we systematically gathered DOIs of scientific articles related to thermoelectric materials, focusing specifically on publications in Elsevier journals. We employed targeted keywords, including “Thermoelectric,” “Seebeck Coefficient,” and “Figure of Merit,” to filter relevant articles from Elsevier’s extensive journal database. Through this process, we compiled approximately 20,000 DOIs pertinent to thermoelectric materials. Using these keywords, we accessed Elsevier’s journal websites and implemented a web-crawling script to automate the retrieval of DOIs.

\subsection{Article Retrieval}
Following above initial collection phase, we used a valid API key provided by Elsevier to submit authenticated API requests, allowing us to download the full-text articles in Extensible Markup Language (XML) format. XML format files include many nested tags and extraneous metadata that necessitate additional processing to extract the relevant content. To address this, we applied a customized text and table parsing tool, designed to extract only the essential information as seen in the PDF version of each article, simplifying it into a plain-text CSV format.

\begin{figure*}[ht]
	\begin{center}
		\includegraphics[width=\textwidth]{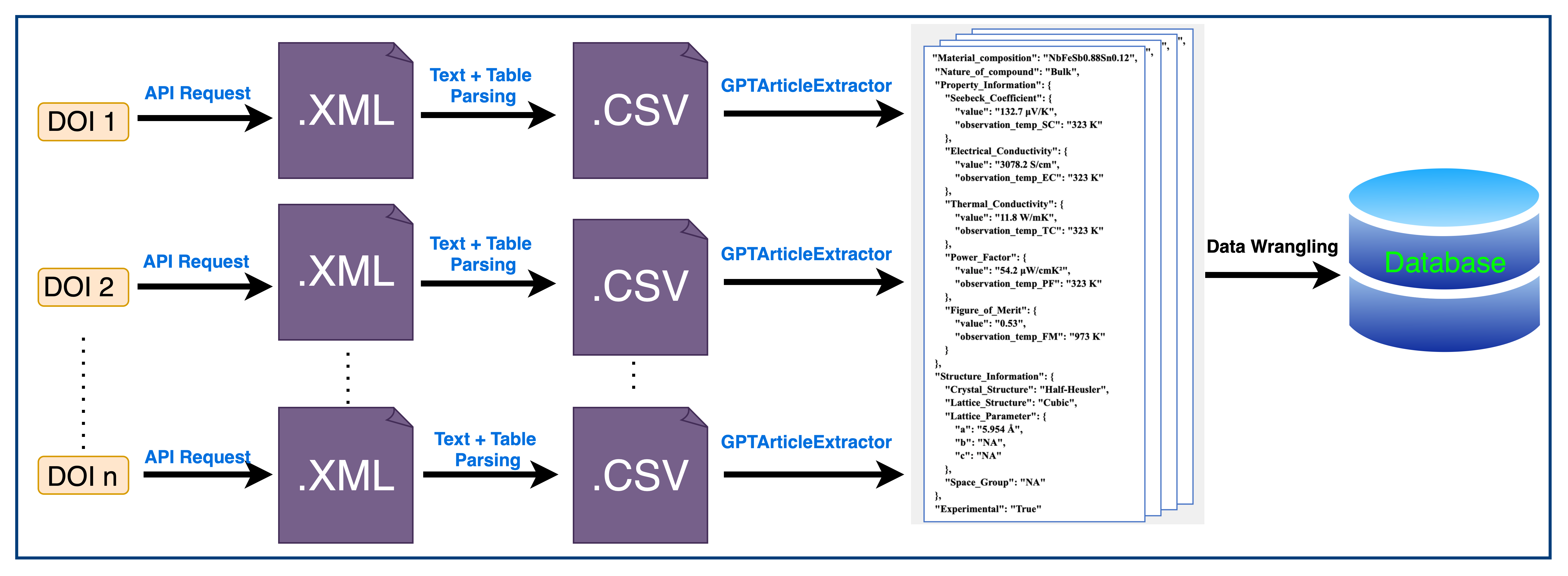}
        \vspace{-30pt}
		\caption{\textbf{Workflow for constructing a database of thermoelectric materials.} Scientific articles, identified by their DOIs, are accessed via API requests and processed as XML files. Using text and table parsers, the XML files are converted into plain-text CSV format. The GPTArticleExtractor, leveraging the GPT-4 model, is then applied to extract relevant information from the plain text and organize it into structured JSON lists. These JSON files are further refined and combined  to create the final database, containing detailed thermoelectric and structural properties of the materials.}
		\label{fig:workflow}
	\end{center}
\end{figure*}

\subsection{Data Extraction and Database Compilation}

The GPTArticleExtractor \cite{zhang2024gptarticleextractor} utilizes the capabilities of modern Large Language Models (LLMs), such as GPT, to extract structured information from large volumes of text. Specifically, it employs the GPT-4 model via the OpenAI API for data extraction. The tool is highly customizable, allowing users to design prompts tailored to extract specific information from articles according to their needs. The extracted data is organized into well-structured JSON files, adhering to the format defined in the input prompt. For articles describing multiple materials with relevant properties, the tool generates a list of JSON objects, each corresponding to a specific material, as demonstrated in Figure \ref{fig:workflow}.

In this study, we applied the GPTArticleExtractor workflow to extract detailed thermoelectric properties of compounds. Prompts were carefully designed based on the methodology outlined in the original article \cite{zhang2024gptarticleextractor}, ensuring the accuracy and relevance of the extracted data. Since the extracted properties were reported in varying units across different articles, we developed a data-cleaning script to standardize all values into uniform units. This final step enabled us to build a clean and comprehensive database with numerous features, ready for further analysis and application.

\section{Results and Discussion}

\subsection{Thermoelectric Materials Database: Structure and Content}

Through efficient extraction and curation of data using large language models, we have constructed a database encompassing 7,123 thermoelectric compounds. This database serves as a central repository of experimentally validated and computationally predicted thermoelectric properties, providing the research community with a standardized and accessible resource.

The database captures critical information about each thermoelectric material, including the chemical formula and the nature of the compound (e.g., bulk, thin film, nanostructured). Accounting for the dimensionality of the materials is crucial, as it can significantly influence their thermoelectric performance.

\begin{figure}[ht]
	\begin{center}
		\includegraphics[width=\linewidth]{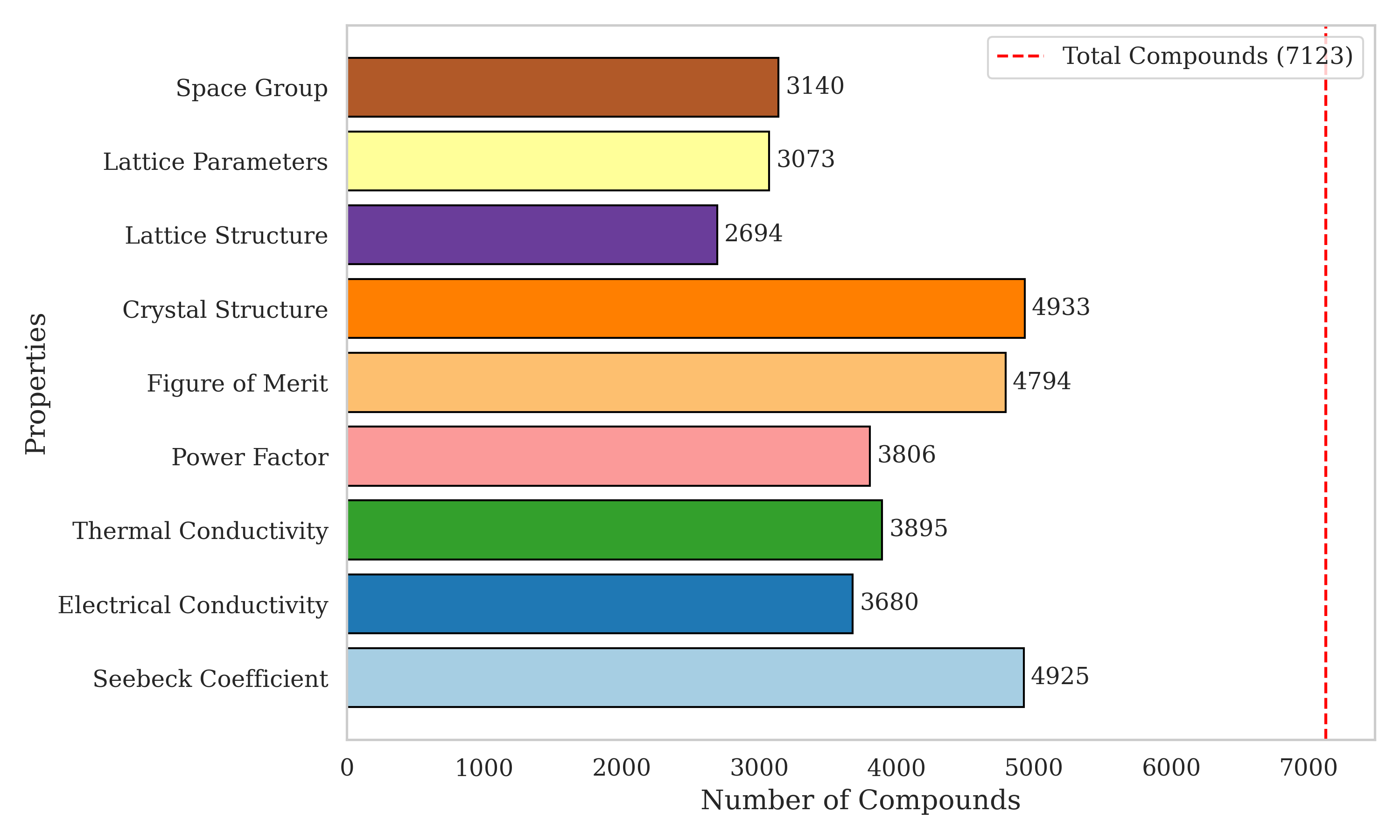}
        \vspace{-20pt}
		\caption{Bar chart plot for the number of compounds for each properties in our database.}
		\label{fig:bar_chart_property}
	\end{center}
\end{figure}

For each entry, the database contains detailed measurements of the key thermoelectric properties, such as Seebeck coefficient, electrical conductivity, thermal conductivity, power factor, and figure of merit (ZT). However, it is important to note that not every publication reports the complete set of properties for a given material. Figure \ref{fig:bar_chart_property} showcases the distribution of data availability for these critical parameters across the database. Additionally, since thermoelectric properties are highly temperature-dependent, we collected the measurement temperature for each property whenever it was provided in the source article.
This allows for the analysis of the temperature-dependent behavior of the materials, which is essential for designing efficient thermoelectric devices.

In addition to the thermoelectric properties data, the database includes structural details of the compounds, such as crystal structure, lattice structure, lattice parameters, and space group.  These structural attributes are essential for exploring structure-property relationships, which play a key role in designing high-performance thermoelectric materials. However, some structural information is incomplete because not all published articles include structural details of the compounds.

The database also includes a feature indicating whether the data originates from experimental work or theoretical calculations, labeled as “True” or “False”. Approximately 66\% of the records in the database come from experimental papers This feature allows users to easily separate experimental data from theoretical data, which supports building machine learning models focused on experimental data.


\begin{table}[ht]
\centering
\caption{Database Content}
\begin{tabular}{|l|l|l|}  
\hline
Features & Type& Unit  \\ 
\hline
Chemical Formula&   String &            \\
Compound Type &    String &               \\
Seebeck Coefficient(S) &  Numeric &  $\mu$V$/$K       \\
$S$ Measurement Temperature & Numeric & K \\
Electrical Conductivity ($\sigma$) & Numeric  &   S$/$m       \\
$\sigma$ Measurement Temperature &   Numeric & K\\
Thermal Conductivity ($\kappa$) & Numeric &    W$/$m.K     \\
$\kappa$ Measurement Temperature  & Numeric & K \\
Power Factor (PF) & Numeric&  $\mu$W$/$m.K$^2$ \\
PF Measurement Temperature &Numeric& K\\
Figure of Merit (ZT) & Numeric&   \\
ZT Measurement Temperature &Numeric& K\\
Crystal Structure & String &   \\
Lattice Structure & String &    \\
Lattice Parameters & Numeric & \AA \\

Space Group & String &   \\
Experimental & Boolean & \\
DOI & String &   \\

\hline
\end{tabular}
\label{table:features}
\end{table}

\subsection{Thermoelectric Property Visualizations and Analysis}
\subsubsection{Seebeck Coefficient}
\begin{figure}[ht]
	\begin{center}
		\includegraphics[width=\linewidth]{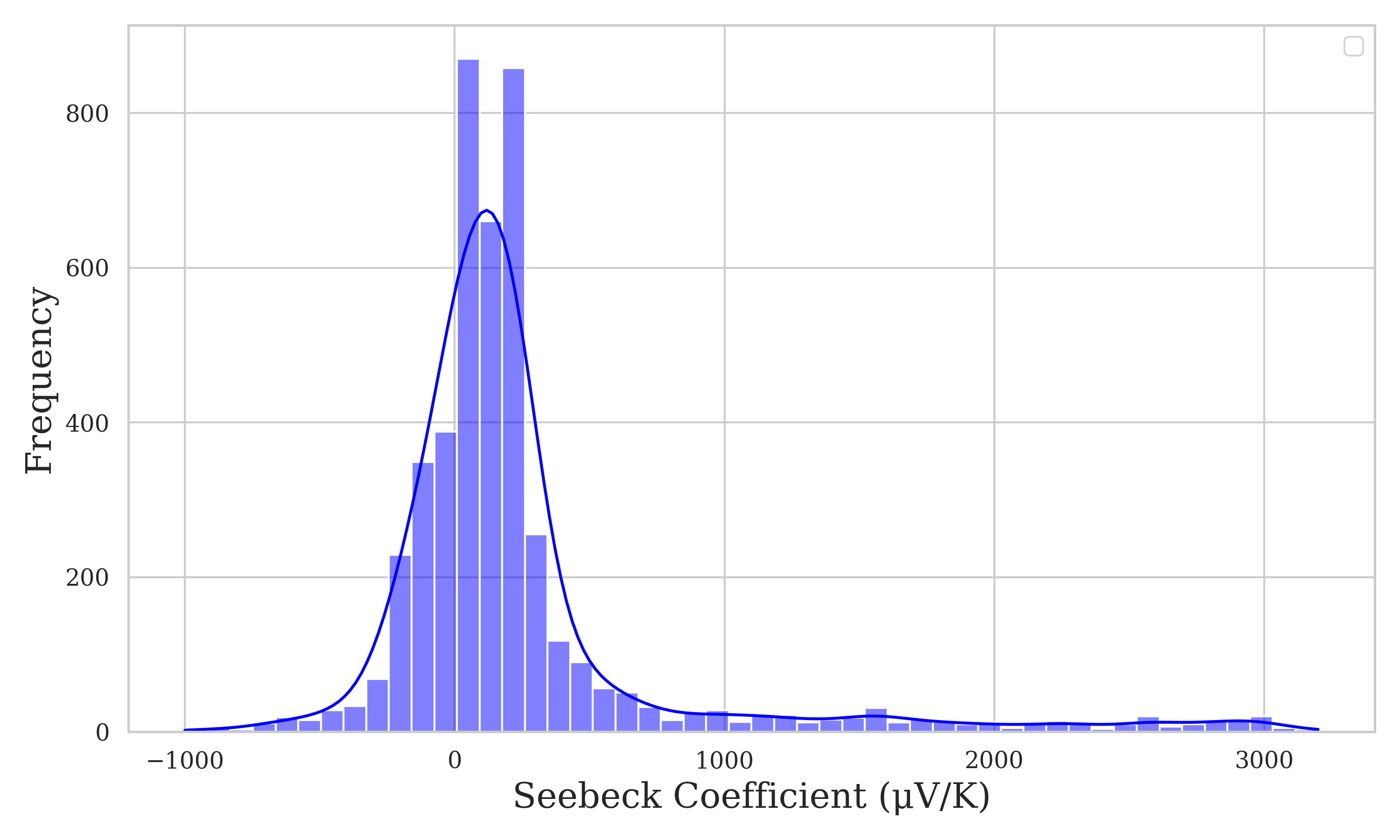}
        \vspace{-20pt}
		\caption{Distribution of Seebeck Coefficient in Database}
		\label{fig:seebeck}
	\end{center}
\end{figure}

The Seebeck coefficient is a fundamental property in thermoelectric materials science, indicating the ability of a material to convert temperature differences into electrical voltage. In Figure \ref{fig:seebeck}, we observe a distribution of Seebeck coefficient values for compounds in our database, with both negative and positive values.
A negative Seebeck coefficient indicates that electrons are the dominant charge carriers in the material, classifying it as an n-type thermoelectric. Conversely, a positive Seebeck coefficient signifies that holes (positive charge carriers) are predominant, making it a p-type thermoelectric. This distinction is critical in thermoelectric applications, as efficient thermoelectric devices require pairing both n-type and p-type materials to establish a thermoelectric couple, which maximizes energy conversion efficiency.

Most values cluster around the low range (-200 $\mu V/K$ to 500 $\mu V/K$), with both n-type and p-type materials represented. However, we also observe some higher values, with a few compounds reaching up to 3000 $\mu V/K$. These rare, high Seebeck coefficients are highly desirable for enhancing thermoelectric performance, although most are derived from computational studies rather than experimental data.

\subsubsection{Electrical Conductivity}
\begin{figure}[ht]
	\begin{center}
		\includegraphics[width=\linewidth]{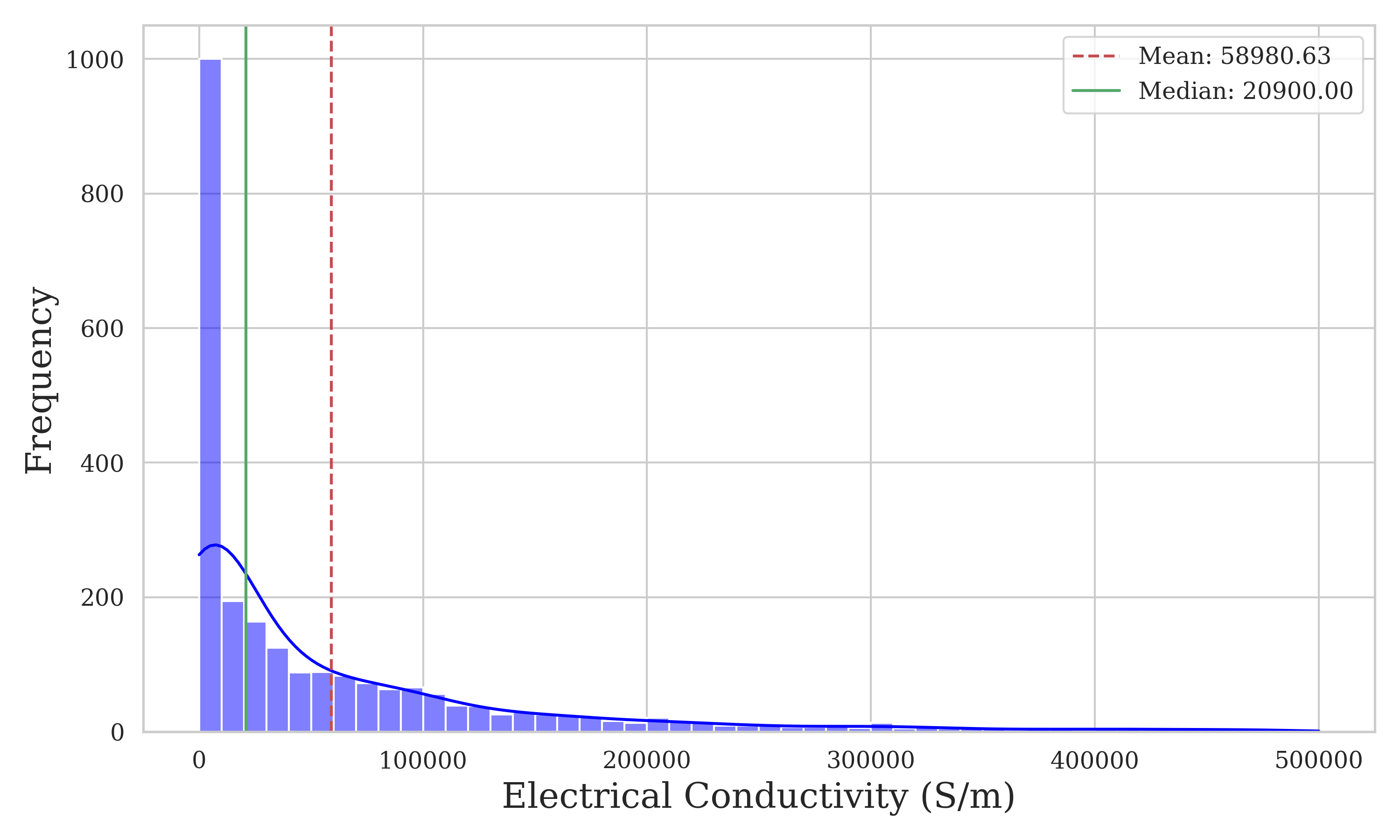}
        \vspace{-20pt}
		\caption{Distribution of Electrical Conductivity in Database}
		\label{fig:electrical_c}
	\end{center}
\end{figure}

Figure \ref{fig:electrical_c} displays the distribution of electrical conductivity values for the thermoelectric materials in our database. The histogram reveals a strong right-skewed distribution, with a majority of values concentrated at lower electrical conductivities, near zero. The mean electrical conductivity is 58,980.63 S/m, indicated by the dashed red line, while the median is lower at 20,900.00 S/m, shown by the solid green line.

A small number of materials exhibit exceptionally high electrical conductivity, reaching up to about 500,000 S/m. Such high-conductivity values are essential for optimizing thermoelectric performance, as higher electrical conductivity helps to reduce resistive losses in thermoelectric devices.  
The distribution demonstrates that the database includes materials with a wide range of conductivity values, including those suitable for advanced thermoelectric applications.

\subsubsection{Thermal Conductivity}
\begin{figure}[ht]
	\begin{center}
		\includegraphics[width=\linewidth]{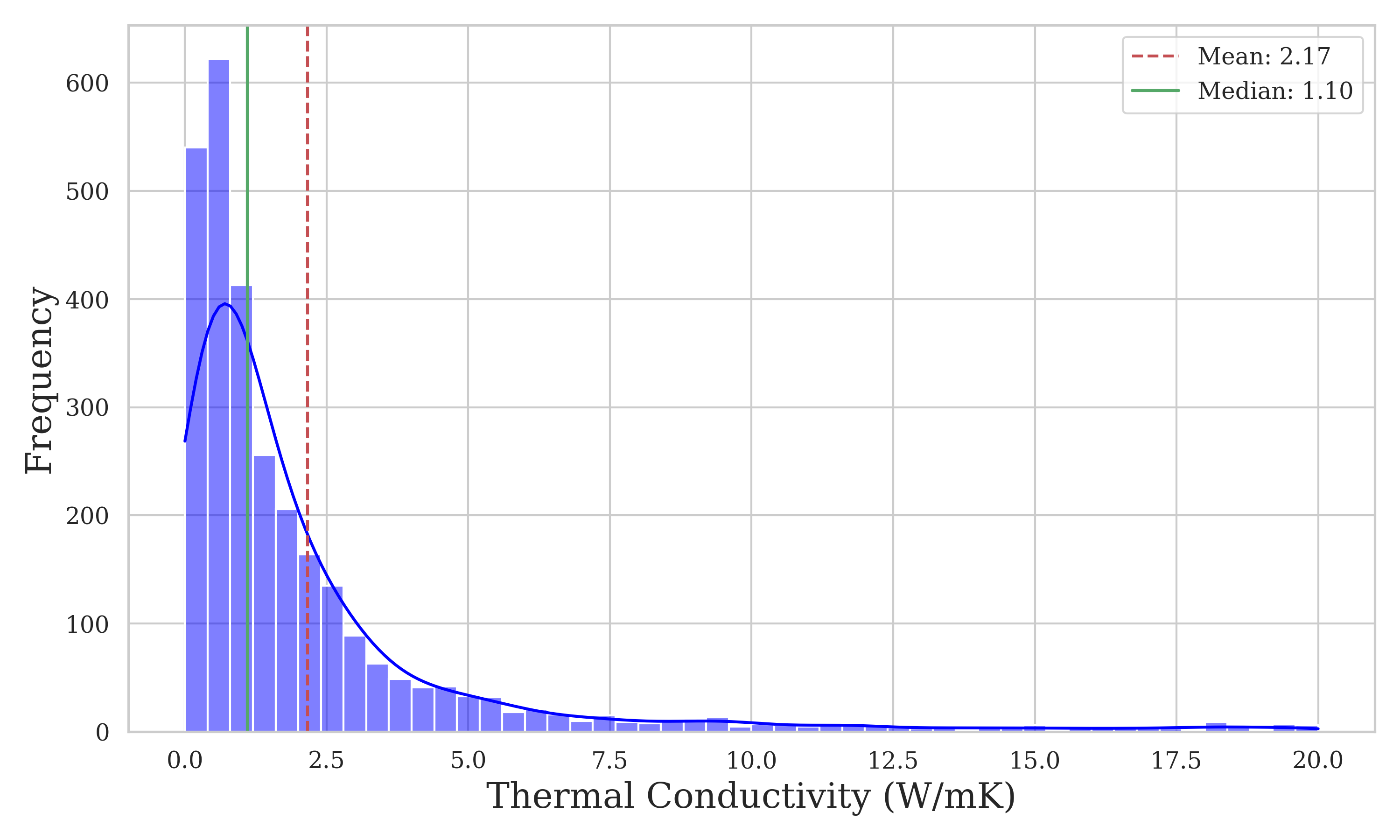}
        \vspace{-20pt}
		\caption{Distribution of Thermal Conductivity in Database}
		\label{fig:thermal_c}
	\end{center}
\end{figure}

Thermal conductivity is a crucial parameter in thermoelectric materials, as lower thermal conductivity generally enhances thermoelectric efficiency by maintaining a temperature gradient across the material. Figure \ref{fig:thermal_c} shows the distribution of thermal conductivity values within our database, with most compounds exhibiting relatively low thermal conductivity.

The histogram demonstrates a strong peak near 1 W/mK, indicating that the majority of materials have low thermal conductivities suitable for thermoelectric applications. The mean thermal conductivity is 2.17 W/mK (dashed red line), while the median is lower, at 1.10 W/mK (solid green line), reflecting a skew towards lower values. A small number of materials extend into higher thermal conductivity ranges, up to around 20 W/mK, although these are rare.

The prevalence of materials with low thermal conductivity aligns with the requirements for effective thermoelectric materials, as reduced thermal conductivity helps to optimize the figure of merit, ZT, in thermoelectric devices. This distribution further highlights the range of thermal conductivities available within our database, offering insights into potential candidates for high-efficiency thermoelectric applications.

\subsubsection{Power Factor}

\begin{figure}[ht]
	\begin{center}
		\includegraphics[width=\linewidth]{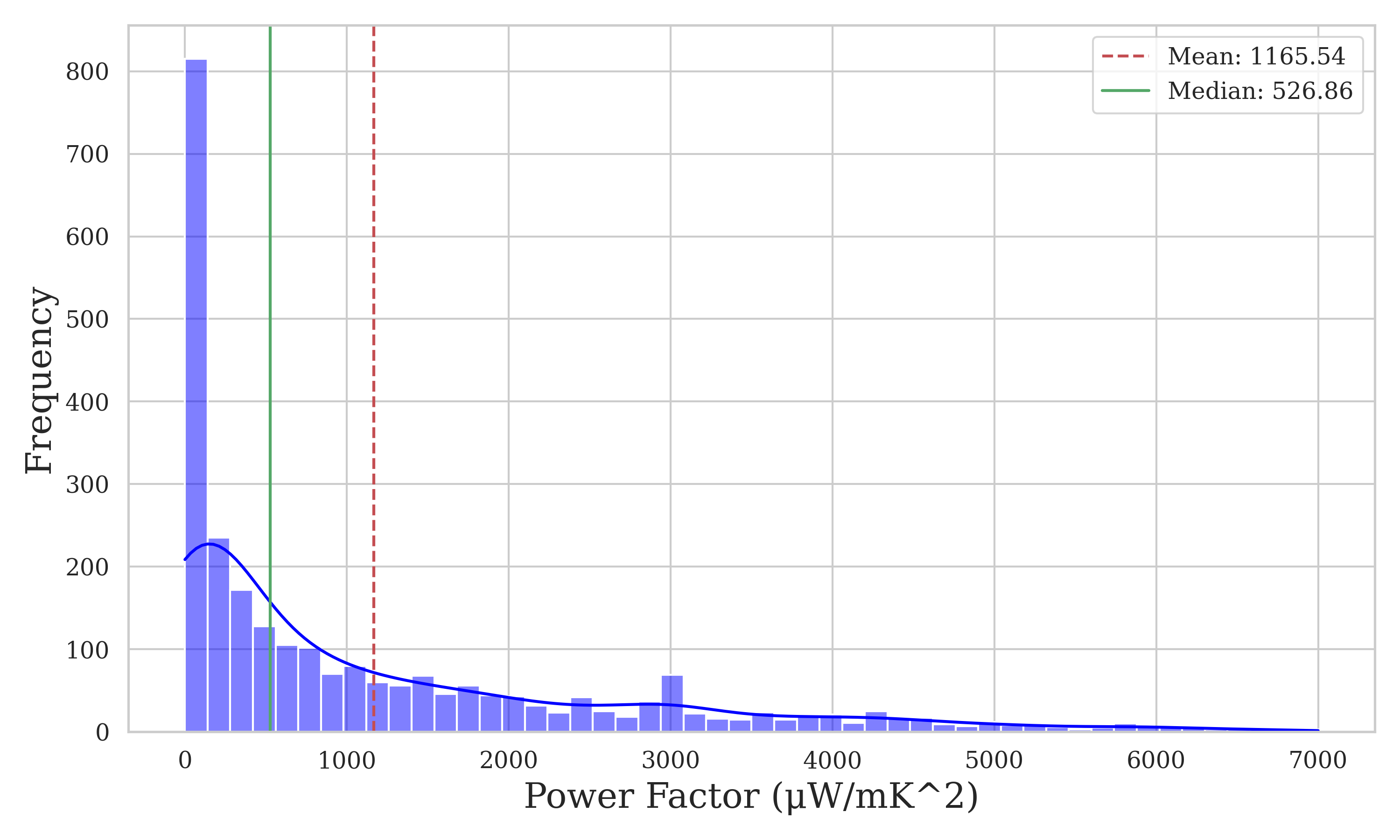}
        \vspace{-20pt}
		\caption{Distribution of Power Factor in Database}
		\label{fig:power_factor}
	\end{center}
\end{figure}

The power factor, defined as $$PF = S^2 \cdot \sigma$$ where S is the Seebeck coefficient and $\sigma$ is the electrical conductivity, is a crucial performance indicator for thermoelectric materials. It reflects a material’s potential for efficient energy conversion by combining these two key properties.\ref{fig:power_factor} illustrates the distribution of power factor values across the thermoelectric compounds in our database.

The histogram reveals a right-skewed distribution with the majority of values clustered at lower power factors, peaking near 0 $\mu W/mK^2$. The mean power factor is 1165.54 $\mu W/mK^2$(dashed red line), while the median is lower at 526.86 $\mu W/mK^2$(solid green line), indicating a skew towards lower values. A small number of materials exhibit exceptionally high power factors, reaching up to approximately 7000 $\mu W/mK^2$, making these materials particularly promising for high-performance thermoelectric applications.

The presence of materials with high power factors underscores the potential of this database to identify candidates suitable for efficient thermoelectric devices.

\subsubsection{Figure of Merit (ZT)}
\begin{figure}[ht]
	\begin{center}
		\includegraphics[width=\linewidth]{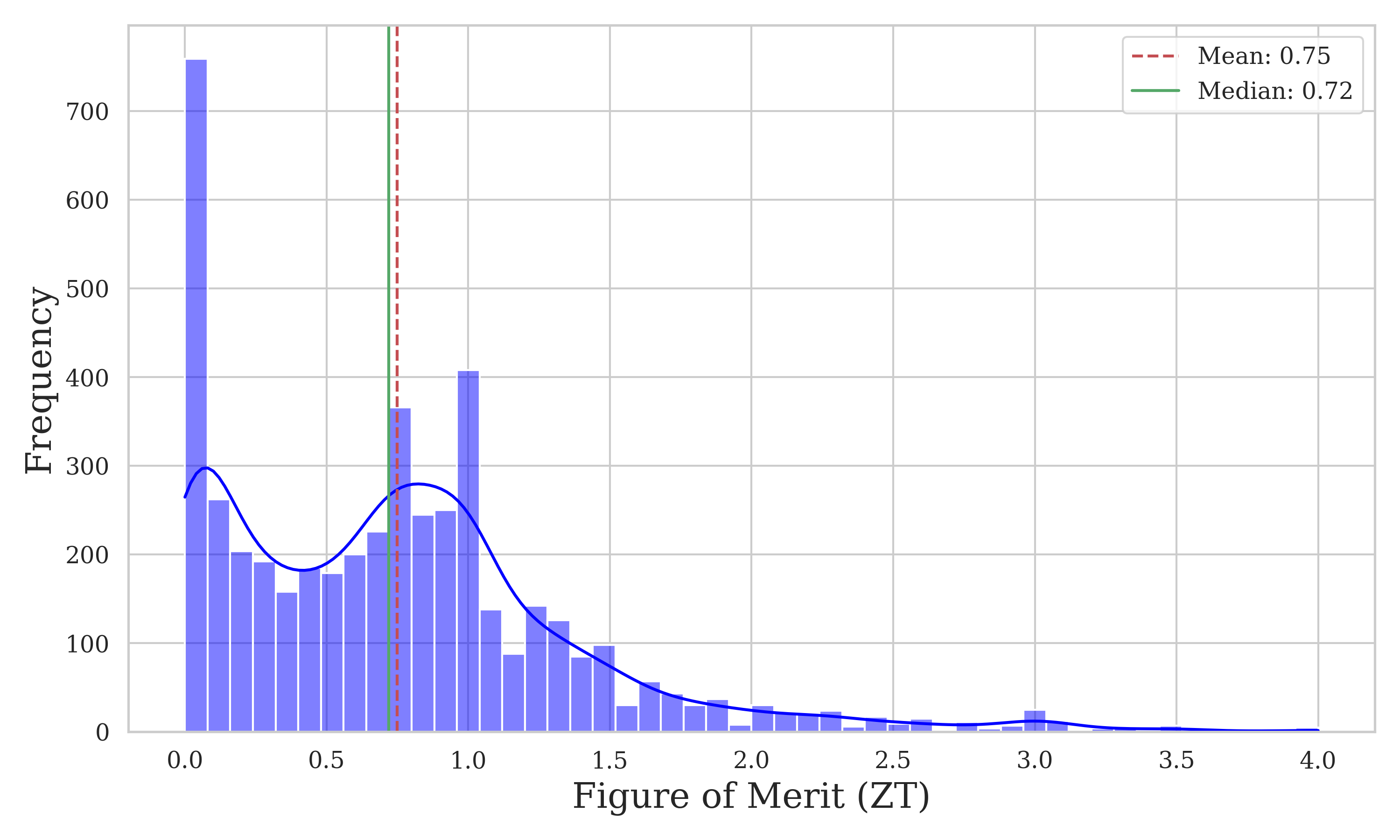}
        \vspace{-20pt}
		\caption{Distribution of Figur of Merit in Database}
		\label{fig:figure_of_merit}
	\end{center}
\end{figure}

The figure of merit, ZT, is a comprehensive metric for evaluating thermoelectric material performance, combining the Seebeck coefficient, electrical conductivity, and thermal conductivity to reflect overall efficiency. Figure \ref{fig:figure_of_merit} shows the distribution of ZT values of thermoelectric compounds within our database.
The histogram reveals that the majority of materials exhibit  ZT  values below 1, with the highest concentration around 0.5 to 1.0, indicating that many materials fall within a moderate performance range. The mean  ZT  value is 0.75 (dashed red line), closely aligned with the median of 0.72 (solid green line), suggesting a relatively balanced distribution around this range. A few materials achieve higher ZT values, extending up to approximately 4.0, although these high-performance compounds are rare.

\section{Conclusions}
We developed a comprehensive database of 7,123 thermoelectric compounds, incorporating key thermoelectric and structural properties using the automated data extraction capabilities of the GPTArticleExtractor. By including structural information, our database is well-suited for future use with advanced methods such as graph neural networks, enabling predictive modeling and material analysis. The database is openly accessible at \href{http://nemad.org}{nemad.org} and is continually updated and expanded. This work provides a valuable resource for data-driven research in sustainable energy, supporting the discovery and analysis of thermoelectric materials and contributing to advancements in the field.

\section*{Acknowledgments}
This work was supported by the Office of Basic Energy Sciences, Division of Materials Sciences and Engineering, U.S. Department of Energy, under Award No. DE-SC0020221. 

\section*{Data availability}
Data will be made available on request.

\bibliographystyle{elsarticle-num} 
\bibliography{reference}

\end{document}